# Injectable Thermochemical Micro-Explosion for Prompt Thrombolysis via Liquid Alkali Metal

Xin Liao[1,5], Yi Hou[2,5], Jie Zhang[2], Bo Wang[3], Minghui Guo[2], Hua Qu[4,*], Wei Rao[2,*], Jing Liu[1,2,*]

1. School of Biomedical Engineering, Tsinghua University, Beijing 100084, China
2. State Key Laboratory of Cryogenic Science and Technology, Technical Institute of Physics and Chemistry, Chinese Academy of Sciences, Beijing 100190, China
3. Institute of Materials Research & Center of Double Helix, Shenzhen International Graduate School, Tsinghua University, Shenzhen 518055, China
4. Xiyuan Hospital of China Academy of Chinese Medical Sciences, Beijing 100091, China
5. These authors contributed equally: Xin Liao, Yi Hou

* Corresponding Author. Email: jliu@mail.ipc.ac.cn; weirao@mail.ipc.ac.cn; hua_qu@yeah.net

## Abstract

Thrombotic vascular diseases contribute to significant global mortality, yet current therapeutic strategies face persistent challenges including bleeding risks, suboptimal efficiency, and procedural complexity. Here, we report a micro-explosive thermochemical thrombolysis (METCT) therapy via injectable liquid alkali metal (LAM) encapsulated in dimethyl silicone (LAM@oil), which enables prompt, efficient and safe vascular recanalization within an ultrafast timeframe (< 90 seconds). This LAM@oil system effectively disrupts thrombus tissue through a synergistic triple-action mechanism: Mechanical micro-explosions forces, alkaline ablation due to highly localized exothermic chemical reactions, and thermal thrombolysis mediated by elevated temperature. Upon thrombolysis completion, the non-toxic reaction byproducts (sodium and potassium ions) exhibit physiologically biocompatible and metabolizable effects. Critically, the LAM@oil demonstrates significantly higher thrombolytic efficacy compared to clinically available thrombolytic drugs (residual thrombus area percent 10.87% ± 7.16% for LAM@oil vs. 80.86% ± 13.32% for urokinase), with no associated bleeding risks. This strategy opens a byproduct-free, cost-effective, and high-efficiency alternative to conventional thrombolytics, holding big potential for clinical translation in acute thrombosis management.

## 1. Introduction

Vascular diseases, particularly thrombus formation, remain a persistent global health challenge, characterized by a high morbidity and mortality rates, especially for cerebral infarction, pulmonary embolism and myocardial infarction (Fig. 1A). These life-threatening conditions impose substantial clinical burdens on healthcare systems worldwide [1,2]. Uncontrolled thrombus formation can lead to devastating ischemic sequelae, including tissue necrosis, multi-organ failure, and permanent disability. These complications can be seen in a wide range of medical conditions, such as myocardial infarction, stroke, disseminated intravascular coagulation, and pulmonary embolism [3,4]. Consequently, timely vascular recanalization, restoration of blood perfusion, and mitigation of ischemic injury are paramount therapeutic goals in thrombotic emergencies. The development of effective antithrombotic strategies thus remains a high-priority research focus.

Thrombotic diseases including myocardial infarction, stroke, and pulmonary embolism, are leading causes of global mortality, accounting for over 30% of annual deaths worldwide. Efficient vascular recanalization is crucial for reducing ischemic injury, yet existing therapies are far from optimal. Current clinical thrombolysis methods can be classified into drug thrombolysis, mechanical-assisted thrombolysis and interventional thrombolysis [5]. The most fundamental and widely used therapy involves the injection of thrombolytic agents to achieve vascular recanalization. The commonly deployed agents include urokinase (UK), tissue plasminogen activator (tPA), streptokinase, reteplase, and Tenecteplase [6,7]. These agents dissolve blood clots by activating the fibrinolytic system. However, this also disrupts normal coagulation, leading to a risk of systemic bleeding-including intracranial hemorrhage in <5% of cases. Hemorrhagic complications range from local hematoma and ecchymosis to systemic manifestations such as hemodynamic instability or intracranial bleeding [8-10]. In addition, thrombolytic agents have ultrashort plasma half-lives (typically a few minutes), resulting in suboptimal therapeutic utilization that necessitates frequent, high-dose systemic administration, which further exacerbate safety concerns [11,12]. Surgical interventions serve as an alternative therapeutic avenue for thrombotic disorders, aiming to disrupt obstructive thrombi and reinstating hemodynamic flow [13-15]. Procedures such as catheter-directed thrombolysis (CDT) involve the precise endovascular navigation of a catheter to the occlusion site for localized administration of thrombolytic agents [16]. Despite its minimally invasive nature, CDT has several limitations, including significant hemorrhagic risks, prolonged treatment durations (often exceeding 6 hours), the need for specialized equipment, and substantially high costs [17]. By virtue of the capacity to prolong systemic circulation duration and minimize off-target biodistribution, therapeutic nanoagents have emerged as one of the most transformative innovations in thrombotic disease management over the last decades [18,19]. Recently, the utilization of micro/nanomotors has facilitated the process of biobarrier-penetrating mechanical therapy, demonstrating distinct advantages as a controllable non-pharmacological modality [20-29]. For instance, Yang et al. engineered heparin-mimetic polymer brush-functionalized magnetic nanorobots that enable safe swarm-based thrombolysis at biological interfaces [30]. Chang et al. developed a novel milli-spinner mechanical thrombectomy technology that achieves ultra-rapid thrombus debulking and high-fidelity revascularization, outperforming aspiration-based thrombectomy approaches [31]. Wang et al. reported biodegradable magnetized biohybrid hydrogel fibers that can evade immune recognition for targeted intracranial tumor therapy [32]. Wan et al. developed platelet membrane-cloaked meso/macroporous silica/platinum nanomotors capable of sequential thrombus-targeting, precision drug delivery, and anticoagulant therapy [33]. Zhang et al. designed a nanoassemble platform that achieved self-indicating and thrombus-targeted accumulation, self-piercing deep clot penetration [34]. These reported nanoplatforms realize outstanding thrombolysis and revascularization through cavitation effect [35] or local thermotherapy [36]. Meanwhile, they also raised important scientific and technological issues to address such as ever-efficient thrombolytic efficacy, less external field support and equipment complexity, as well as maintaining long-term therapeutic effect and biological safety. In this sense, developing simpler, safer, more cost-effective, and highly efficient therapeutic paradigms is still urgently needed in the area.

To address the above challenges, here we propose and demonstrate a micro-explosive thermochemical thrombolysis (METCT) strategy that enables rapid and efficient thrombus dissolution in vivo within exceptionally short timeframes. The principle is based on the synergistic interplay of thermothrombolysis, hydroxyl ions ablation and microbubbles collapse through the injection of our engineered liquid alkali metal (LAM) particles as the source for the reaction with water (Fig. 1B). LAM used in this study is the eutectic NaK alloy (comprising 77.8 wt% K and 22.2 wt% Na), with a metling point of 12.6 °C). LAM exhibits exceptional chemical reactivity, manifesting vigorous explosive reactions upon aqueous contact [37-41]. Notably, NaK alloy has long been an under-explored material in the biomedical field, with few studies investigating its therapeutic potential. Our group had ever proposed its application as a potent self-heating seed for minimally invasive thermochemotherapy in tumor treatment [42,43], marking the systematic validation of NaK's feasibility and efficacy in biomedicine. This prior work focused on verifying the core therapeutic capacity of NaK, the intense thermochemical ablation generated by alkali metal-water reactions releases substantial thermal energy, and NaK administration induced remarkable anti-tumor efficacy in murine mammary diseased models [44]. Alloy-induced ablation therapy elevates localized temperatures to 85 °C at the ablation site, achieving tumor suppression rates of 88.5% in early-stage cancers and 67.6% in developing lesions [45]. These outputs conclusively demonstrated that NaK possesses inherent therapeutic properties that warrant further exploration for broader biomedical applications beyond tumor therapy. Translating NaK's powerful thermochemical properties to thrombus therapy faces challenges, including narrowing the safety window to avoid damaging healthy vascular tissue while ensuring thrombolytic efficacy, developing an injectable formulation suitable for intravascular delivery, establishing clinically relevant thrombotic disease models, and elucidating the interaction mechanisms between NaK and vascular/tissue components. In addition, the inherent high reactivity of alkali metals poses significant ignition and explosion hazards, which is a major barrier for LAM-based biomedical applications. To achieve a safer and controlled thermal and chemical reaction, we stabilized LAM within an inert dimethylsilicone matrix via ultrasonication-mediated emulsification, effectively constraining its inherent violent reactivity. The resultant oil-encapsulated LAM (LAM@oil) exhibits superior injectability for therapeutic delivery. Especially, this configuration forms a protective interfacial film on NaK particles that moderates water-contact kinetics, thereby precisely regulating heat release and reaction duration. Critically, the dimethylsilicone imparts inherent antithrombogenicity by minimizing plasma protein and cellular interactions, specifically attenuating fibrinogen adsorption and platelet adhesion [46]. Consequently, the LAM@oil precludes the occurrence of secondary thrombosis in vivo. As illustrated in Fig. 1C, within the thrombotic microenvironment, LAM particles in the mixture react with interstitial water, generating dense hydrogen microbubbles and localized hyperthermia. The reaction kinetics are regulated by concentration-dependent modulation of LAM and oil blending ratios. Subsequent collapse of hydrogen microbubbles under thermal excitation initiates cavitation effects, producing high-energy microjets that fracture fibrin networks and disrupt thrombus integrity. Continuous microbubble generation, coupled with iterative micro-explosions, progressively degrades the thrombotic structure, thereby accelerating thrombolysis. At thrombotic sites, LAM@oil simultaneously initiates vigorous thermochemical reactions and generates hydroxide ions, establishing an alkaline microenvironment that potentiates therapeutic ablation efficacy through protein hydrolysis. Concomitant temperature elevation induces disruption of non-covalent interactions within thrombotic proteins, thereby facilitating structural destabilization and mediating thermothrombolysis. Notably, the resultant sodium and potassium ions serve as green and essential electrolytes that play a critical role in maintaining cardiovascular homeostasis. Moreover, the controlled release of these ions beneficially modulates cellular energy metabolism pathways.

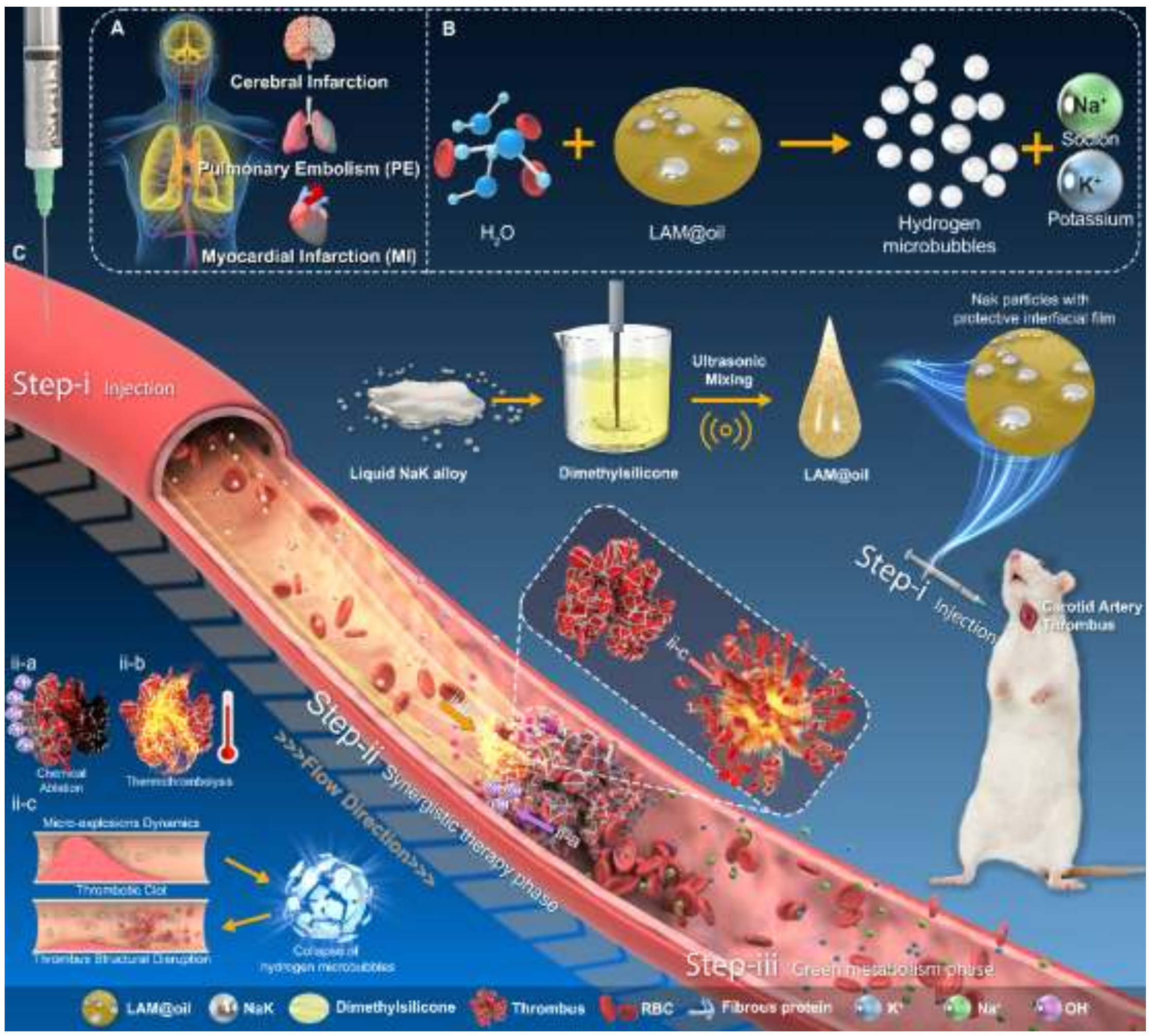

**Fig. 1 Principle of liquid alkali metal (LAM) enabled micro-explosive thermochemical thrombolysis (METCT) strategy.** A, Typical thrombotic diseases: cerebral infarction, pulmonary embolism (PE), myocardial infarction (MI). B, Mechanism of LAM@oil mediated METCT therapy. C, Schematic illustration of the fabrication, biosafety, therapeutic mechanism, and applications of the LAM@oil thrombolytic strategy. Step-i: Intracarotid administration in a rat thrombus model: The LAM@oil mixture, generated by ultrasonication-mediated emulsification of dimethylsilicone and LAM, was injected at the proximal site of the thrombus. Step-ii: Synergistic therapy phase: Accompanying therapeutic reactions include ii-a) Chemical ablation ($OH^-$ as a transient therapeutic agent), ii-b) Thermothrombolysis, and ii-c) Micro-explosion (hydrogen microbubble collapse) effect. Step-iii: Green metabolism phase: Thrombus dissolution with the release of biocompatible byproducts ($Na^+$ and $K^+$).

To our knowledge, this work presents the first ever engineered thrombolytic agent based on LAM@oil, establishing a potential injectable strategy and providing mechanistic insights into multimodal synergistic thrombolysis. Distinct from conventional approaches, this system integrates: (i) Micro-explosive disruption: hydrogen microbubble implosion generates cavitational microjets to fracture fibrin matrices. (ii) Alkaline ablation: localized $OH^-$ ions hydrolyze thrombus proteins. (iii)

Thermal thrombolysis: hyperthermia denatures coagulation factors. Therefore, LAM@oil achieves exceptional thrombolytic efficiency with ultra-low dosage requirements and shortened treatment timelines. Critically, the LAM@oil demonstrates a favorable biosafety profile in whole-blood environments, producing benign biocompatible byproducts and eliciting no significant organ toxicity or inflammatory pathology. This work paves out an accessible, rational, and economically viable therapeutic strategy with scalable clinical deployment potential and transferable application pathways.

## 2. Results

### 2.1 Fabrication and thermochemical characterization of LAM@oil

To achieve efficient thrombus dissolution within a short timeframe, we engineered and prepared an injectable thrombolytic agent with biocompatibility, green feature and safety. The LAM (NaK) exists in a liquid state at room temperature with excellent fluidity, allowing it to be dispersed into micro/nanoscale particles via ultrasonication. The LAM (room-temperature melting point: −12.65 °C) was uniformly dispersed as microdroplets in dimethylsilicone using probe sonication. Dimethylsilicone provides critical stabilization through its low surface tension, high chemical stability, and non-toxicity, forming protective interfacial films that prevent metal coalescence. As previously noted, a primary challenge in handling LAM is the risk of ignition or explosion due to its high reactivity (Fig. 2A). As demonstrated in Fig. 2B, the instant contact of trace LAM (≤1 μL) with phosphate-buffered saline (PBS) triggered violent explosions with visible sparks (Fig. 2B(c), (d)). Therefore, such uncontrolled exothermic reactions preclude further in vivo application.

To facilitate its biomedical application, we developed safer handling techniques for LAM. Following the technical route in Fig. 2C, silver-gray liquid NaK was injected into transparent dimethylsilicone, and the mixture was sonicated to a homogeneous dark gray NaK dimethylsilicone mixture (LAM@oil) with stabilized reactivity suitable for thrombolytic deployment (Fig. 2C).

To investigate the reaction kinetics of the fabricated LAM@oil at varying concentrations, we prepared formulations with distinct volume ratios of LAM to oil. As shown in Fig. 2D(a), 10 μL of LAM was injected into 20 μL, 40 μL, or 80 μL of dimethylsilicone (oil), followed by sonication to obtain gray LAM@oil mixtures of graded concentrations (Fig. 2D(b)). Notably, the resulting mixtures exhibited injectable viscosity, enabling easy loading into syringes for clinical handling (Fig. 2D(c)). To simulate physiological conditions in biological systems, we employed PBS as the reaction medium. When LAM@oil of different concentrations was added to the PBS solution, we observed a gradual reaction between the active component (NaK) in the system and the PBS solution over time. This reaction was visually evidenced by the distinct color transition of the LAM@oil from gray to transparent. The thrombolysis reaction rate exhibited a concentration-dependent relationship, with complete reaction typically achieved within specific timeframes. Specifically, LAM@oil with higher dimethylsilicone dilution (10 μL LAM-40 μL oil, 10 μL LAM-80 μL oil) completed the reaction within 7 min, exhibiting full transparency with sparse residual bubbles. At the same time, the highest-concentration formulation (10 μL LAM-20 μL oil) also completed the reaction in 29 minutes. Importantly, the reaction proceeded in a controlled and safe manner, with no violent explosions or sparks generation. These results demonstrate the excellent stability and high operational safety of the proposed LAM@oil system, which successfully mitigated the inherent high reactivity of pure LAM (Fig. 2E). Furthermore, to directly verify the occurrence of micro-explosions at the microscale, we performed detailed experiments by coupling a high-speed macro-lens camera with a microscopic imaging system. High-speed footage clearly captured the generation of dense hydrogen microbubbles at the LAM@oil-PBS interface within 180 s after administration. Subsequent collapse of these microbubbles generated microscale jet streams, which appeared as transient bright streaks in the imaging data, consistent with the cavitation effects characteristic of micro-explosions.

To investigate the thermal characterization of LAM@oil, various LAM volumes (5 μL, 10 μL, and 20 μL) combined with different oil volumes (20 μL, 40 μL, 80 μL, and 100 μL) were prepared. Fig. 2F illustrated the time-dependent temperature profiles of LAM@oil, with the same volume of LAM and different volumes of oil. Upon injection, these mixtures induced a rapid and substantial temperature increase in the target reaction zone, elevating from an initial temperature (28.59 °C) to 56.43 °C ($\Delta T$ = 27.84 °C) within a remarkably short duration of 22~34 seconds. This pronounced thermal effect demonstrated significant potential for enhancing thrombus dissolution. It was shown that for the same LAM volume (20 μL), varying oil proportions resulted in comparable temperature escalation rates and peak temperature amplitudes in the target reaction zone (Fig. 2F). Thermal analysis revealed that the most substantial temperature increase occurred within the first minute of reaction for all tested LAM@oil ratios, demonstrating the rapid onset of the thrombolytic agent and its ability to quickly reach therapeutic temperatures for thermal thrombolysis. For a fixed LAM volume (20 μL), comparative evaluation of three distinct LAM-to-oil ratios showed that higher LAM concentrations (e.g., 20 μL LAM@40 μL oil, 1:2 ratio) produced sharper temperature peaks while maintaining similar duration profiles (Fig. 2G). When the oil volume was constant (80 μL), comparative thermal analysis of three LAM-to-oil ratios revealed that under higher LAM concentrations (e.g., 20 μL LAM@80 μL oil mixture, 1:4 ratio), the target reaction zone of the action corresponds to a higher temperature elevation, a longer duration of high temperature and a thinner peak temperature change (Fig. 2G). Our systematic investigation of temperature profiles revealed that with fixed LAM volumes, the maximum temperatures achieved by LAM@oil remained remarkably consistent across different oil proportions. In addition, peak temperatures exhibited strong LAM concentration-dependent, reaching 38.35 ± 0.37 °C (5 μL LAM), 43.78 ± 0.08 °C (10 μL LAM), and 56.14 ± 0.35 °C (20 μL LAM) (Fig. 2H). Notably, under conditions of fixed oil volume, the target zone's maximum temperature demonstrated a positive correlation with increasing LAM-to-oil ratios, confirming the dominant role of LAM content in thermal regulation. In contrast, the pure LAM exhibited a sharp and rapid increase in temperature, exceeding 200 °C, which was extremely uncontrollable.

Collectively, the temperature-time profiles of thrombolytic LAM-oil mixtures with different ratios exhibited generally similar patterns, with no statistically significant differences in peak temperatures. As the LAM-to-oil ratio increased, the rise and fall of the curve became markedly steeper, indicating that the LAM-oil mixture can effectively regulate both the reaction kinetics between LAM and water in tissues and the corresponding rate of heat release. Regarding the dosage of LAM, excessive temperature increases may cause tissue burns due to violent reactions, while insufficient temperature elevation fails to achieve the therapeutic effect required for thermal thrombolysis. Therefore, appropriate ratios and dosages should be carefully selected based on specific clinical scenarios and tissue characteristics.

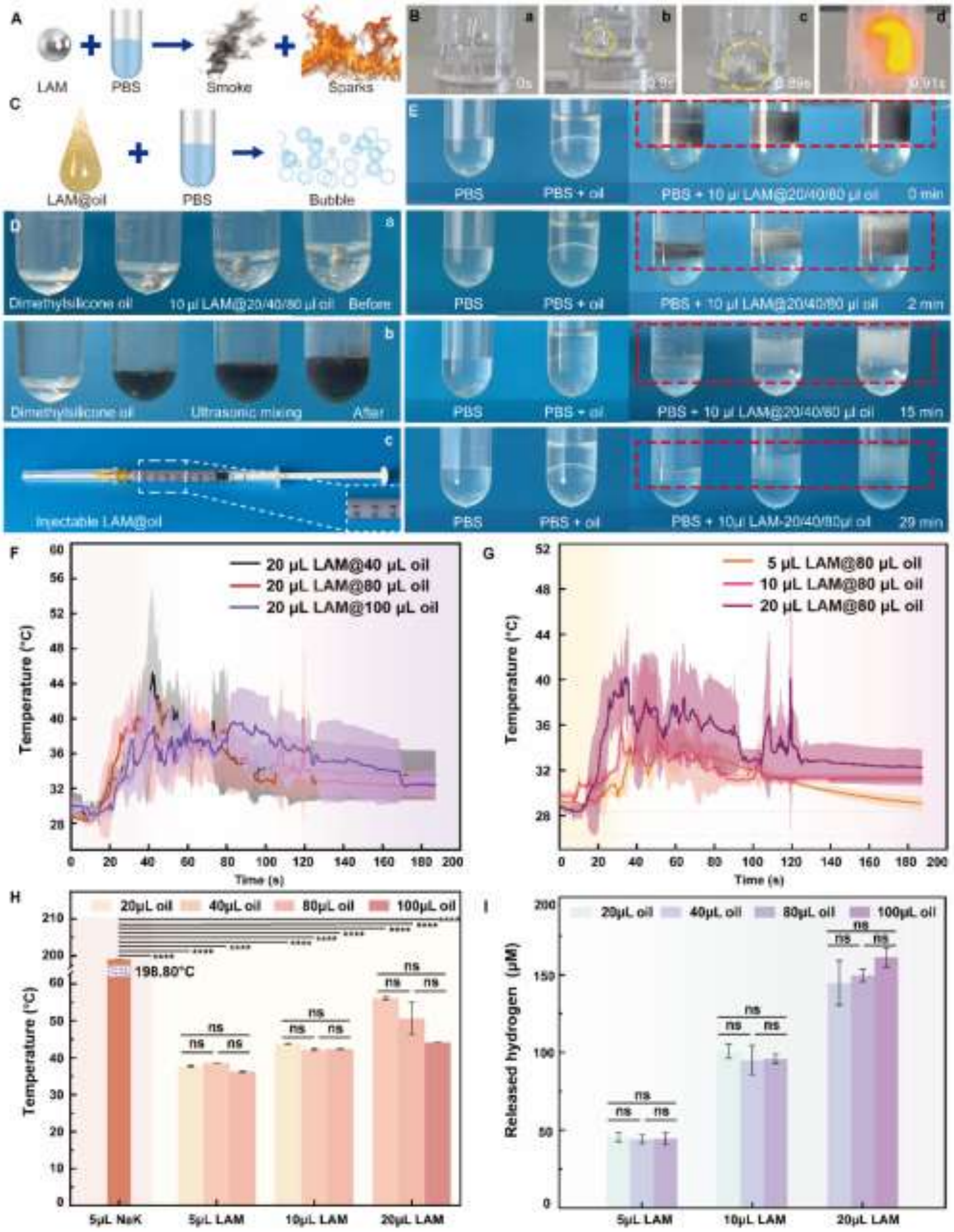

**Fig. 2 Preparation and thermochemical characterization of LAM@oil.** A, Schematic of untreated LAM reaction dynamics in PBS solution. B, Typical reaction images of untreated LAM (1 μL) with PBS solution. A large amount of instantaneously generated explosive smoke and vigorous sparks are concomitantly produced. C, Schematic of LAM@oil reaction dynamics in PBS solution. D, Preparation of LAM@oil at varying concentrations. a (before) and b (after), Dimethylsilicone (control), 10 μL LAM@20 μL oil, 10 μL LAM@40 μL oil, 10 μL LAM@80 μL oil (*left to right*). c, Injectability demonstration of LAM@oil. E, Controlled reaction kinetics of LAM@oil with PBS across compositional gradients. PBS (control), oil, 10 μL LAM@20 μL oil, 10 μL LAM@40 μL oil, 10 μL LAM@80 μL oil (*left to right*). F, Temperature-time profiles for fixed LAM dosage (20 μL) with varied oil volumes (40, 80, and 100 μL). G, Temperature-time profiles for fixed oil volume (80 μL) with varied LAM volumes (5, 10, and 20 μL). H, Peak temperatures at targeted reaction sites across LAM@oil with distinct compositional ratios. Data are presented as mean ± SD (n=3, independent experiments). One-way ANOVA with Tukey's multiple comparisons

test was used for the analysis of data. The n.s. represent no significance and ****$p < 0.0001$. I, Hydrogen evolution performance of LAM@oil with distinct compositional ratios. Data are presented as mean ± SD (n = 3, independent experiments). One-way ANOVA with Tukey's multiple comparisons test was used for the analysis of data. The ns represents no significance.

The reaction between LAM and water generates hydrogen gas, as described by the following chemical equation (1). To comprehensively evaluate the chemical gas formation, we systematically investigated the hydrogen evolution properties of LAM@oil. As shown in Fig. 2I, for a given LAM volume, hydrogen production remained consistent across different oil proportions. For a fixed oil volume, hydrogen evolution showed direct proportionality to LAM content, confirming that LAM concentration rather than oil is the primary determinant of gas generation. Furthermore, the stable and homogeneous LAM@oil system can be effectively sterilized via UV irradiation for a minimum duration of 10 minutes prior to subsequent experimental procedures.

$$2Na_{0.222}K_{0.778(l)} + 2H_2O_{(l)} = 0.444NaOH_{(aq)} + 1.556KOH_{(aq)} + H_{2(g)} \quad (1)$$

## 2.2 Evaluation of the LAM@oil mediated heat and mass transfer effect

To systematically evaluate the heat and mass transfer effect of LAM@oil during thrombolysis, numerical simulations were performed to quantify the temperature response and ion concentration of blood. As illustrated in Fig. 3A, a cylindrical computational model was established to mimic the in vivo thrombolytic scenario, where LAM and LAM@oil were administered intravascularly and interacted with pre-formed thrombi. This model was constructed based on the geometric and physiological parameters derived from in vitro experiments and in vivo animal studies, ensuring its relevance to real biological systems. All simulations were performed using Comsol Multiphysics (a multi-field coupled simulation platform), with the Pennes bioheat transfer equation and Fick's second law of diffusion integrated to describe heat propagation and ion transport, respectively. Fig. 3B depicted time-dependent temperature profiles at the thrombus center (coordinate (0,0)) for both LAM and LAM@oil groups. It was evident that the temperature of the thrombus increased dynamically with the progression of the reaction, reaching its maximum value within 20 seconds. This indicated that the reaction became more intense within this time frame, which was consistent with the rapid reaction kinetics of LAM with water. The peak temperature in the LAM@oil group was 56 °C, whereas the unencapsulated group exhibited an extremely high peak temperature of 203 °C. These simulation results were in good agreement with the experimental data observed in the actual measurements (Fig. 2H), validating the reliability of the computational model. To further assess the spatial distribution of thermal effects, temperature profiles were analyzed at three additional representative locations within the thrombus:(0, 0.001), (0.001, 0.001) and (0.003, 0.001) (Figs. 3C-3E, respectively). It was evident that the temperature response demonstrated analogous increasing trends at the initial time with a slight right shift. Notably, the unencapsulated LAM group displayed maximum temperatures exceeding 100 °C across all measured positions, which far exceeds the thermal tolerance threshold of vascular tissues In contrast, the maximum temperature in the LAM@oil group was only 56 °C and nearly 42 °C at the peripheral regions of the thrombus, within the safe range for biological tissues while still enabling thermal denaturation of fibrin (a key component of thrombi). This controlled thermal release is attributed to the dimethylsilicone shell, which acts as a physical barrier to moderate the contact rate between LAM and water, thereby mitigating the violent exothermic reaction of pure LAM. Fig. 3F illustrates the 2D temperature distribution within the thrombus and surrounding blood vessels for both groups. The LAM group exhibited extensive thermal diffusion, with high-temperature regions (>100 °C) spreading beyond the thrombus boundary and into the adjacent vascular tissue-posing a severe risk of thermal injury. In contrast, the LAM@oil group showed highly localized heat accumulation, confined primarily to the thrombus region. This spatial confinement of thermal effects is critical for minimizing off-target tissue damage while ensuring effective thermal thrombolysis.

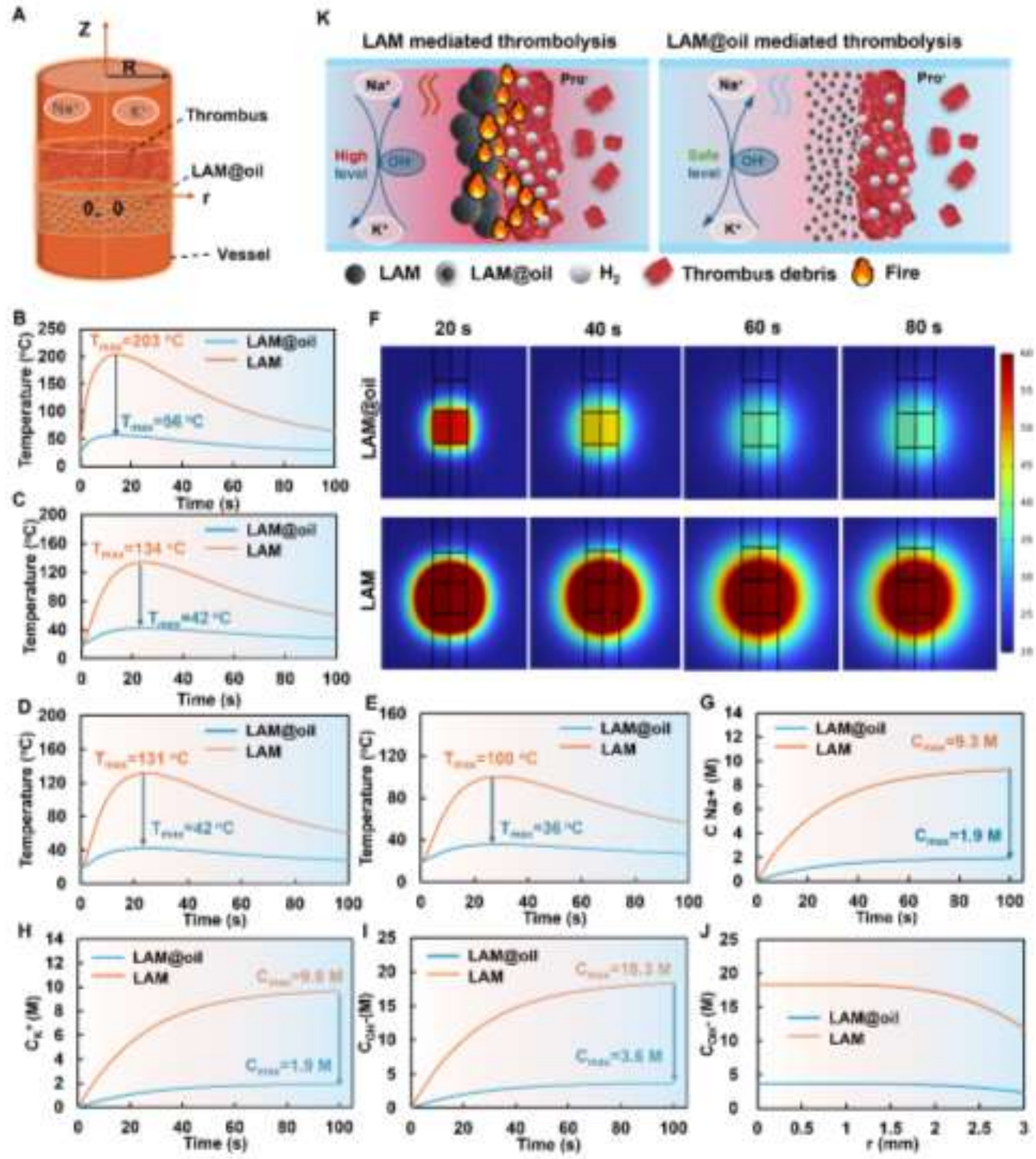

**Fig. 3 Temperature and ion concentration response in LAM@oil mediated thrombolysis model.** A, Cylindrical bioheat and mass transfer geometry model of LAM@oil mediated thrombolysis. B, Temperature response curves of the thrombus center (0,0) of LAM@oil and LAM-mediated thrombolysis. Temperature response curves of the thrombus at coordinate C, (0,0.001), D, (0.001, 0.001), E, (0.003, 0.001). F, Temperature distribution at the location of the thrombus in LAM@oil and LAM-mediated thrombolysis groups. Ion concentration response at the thrombus center (0,0) of G, $Na^+$, H, $K^+$, I, $OH^-$. J, Comparison of the alterations in the concentration of $OH^-$ across various locations along the r-direction of LAM@oil and LAM mediated thrombolysis. K, Mechanism of thrombolysis induced by elevated temperature and pH of LAM@oil and LAM-mediated thrombolysis.

In addition to the thermal behavior, the concentration dynamics of sodium ($Na^+$), potassium ($K^+$), and hydroxide ($OH^-$) ions at the thrombus center were quantified (Figs. 3G–I). For both LAM and LAM@oil groups, ion concentrations increased gradually as the reaction progressed, accompanied by a concurrent decrease in increasing rates. This phenomenon was consistent with the observed trend in reaction temperature. However, the concentrations of sodium, potassium, and hydroxide ions in the LAM group exhibited a fivefold increase relative to that of the LAM@oil reaction, which was also substantially greater than the system's maximum tolerance level. To evaluate the spatial uniformity of ion distribution, $OH^-$ concentration profiles were analyzed along the radial direction (r-direction) of the thrombus (Fig. 3J). The LAM@oil group showed a decrease in $OH^-$ concentration, whereas the LAM group exhibited sharply elevated $OH^-$ concentrations across the entire thrombus and adjacent regions, highlighting the risk of excessive alkaline ablation. Collectively, these simulation results demonstrate that the LAM@oil system achieves a safer spatiotemporal regulation of both thermal and chemical effects compared to unencapsulated LAM (Fig. 3K). The dimethylsilicone shell not only suppresses the violent exothermic reaction of LAM but also controls the release rate of heat, ensuring that the thermal effects are localized to the thrombus while avoiding high temperature injury. This controlled heat and mass transfer behavior is a key prerequisite for the in vivo safety and efficacy of the LAM@oil system.

## 2.3 Quantification of thrombolytic efficacy *in vitro*

The successful synthesis of the thrombolytic LAM@oil and its optimized functional properties motivated further investigation into its thrombus-dissolving efficacy. We first evaluated its thrombolytic ability in vitro. As schematically detailed in Fig. 4A, the reaction mechanism involves LAM reacting with tissue-bound water, generating hydrogen microbubbles ($H_2$), sodium ions ($Na^+$), potassium ions ($K^+$), hydroxide ions ($OH^-$), and substantial exothermic energy release. In the absence of encapsulation, trace amounts of LAM (1 μL) reacted instantaneously with both the thrombus and ambient moisture. Within 2 seconds, substantial thrombus dissolution was observed, accompanied by a visible darkening of the PBS solution (Fig. 4B(b) and (c)). By 1 minute, the supernatant exhibited a significant color change, and thrombus volume was markedly reduced, the reaction was basically over (Fig. 4B(e)). Complete thrombus dissolution occurred within 9 minutes (Fig. 4B(f)). We also measured the thrombolytic capacity of 1μL non-encapsulated LAM, the post-treatment residual thrombus weight curve demonstrated approximately 15.0% thrombolysis rate. Furthermore, untreated thrombi present dense homogeneous structures featuring intact surface fibrin networks and tightly aggregated erythrocytes. Thrombi treated with LAM exhibit distinct therapeutic traces, characterized by significant volume contraction and worm-eaten defects along the margins, indicating localized disintegration of fibrin networks with transformation from dense structures to loose flocculent aggregates. The reaction interface currently presented a charred black appearance, attesting to the destructive effect of thermochemical ablation on protein polymer matrices. The significant morphological alterations in thrombi-manifested as dense-to-loose transformation, intact-to-defect transition, and homogeneous-to-cavitated conversion demonstrated high concordance with micro-explosive damage induced by hydrogen cavitation as previously described. Collectively, these changes visually confirmed that LAM achieved efficient in vitro thrombus destruction through a triple synergistic mechanism, providing direct morphological evidence of successful thrombolysis.

We then evaluated the thrombolytic efficacy of the LAM@oil. Unlike pure LAM, which reacted violently with water accompanied by sparking, the encapsulated mixture modulated the reaction rate between LAM and water molecules when applied to the thrombus. Within this mixture, the LAM was dispersed into numerous microscopic particles. The hydrogen bubbles generated by these particles undergo localized micro-explosions due to the elevated temperature. The resulting microjet exerted a mechanical force that disrupted the thrombus structure. As illustrated in Fig. 4C, upon the injection of LAM@oil into the thrombus model, high-density microbubble clusters generated during the therapeutic reaction can be observed within 6 minutes. The mechanical forces from bubble rupture caused effective disruption of thrombotic tissue (Fig. 4C(c)). By 27 minutes, the reaction was substantially complete, accompanied by visible supernatant turbidity and an apparent reduction in thrombus volume, demonstrating its thrombolytic effect (Fig. 4C(e)). Fig. 4D further evaluated the in vitro thrombolytic efficacy of the LAM@oil. Color changes in the supernatant and reduced thrombus volume were observed in all three experimental groups compared to the control. Untreated LAM demonstrated notable thrombolytic efficacy. Fig. 4D(II) revealed a significant increase in fibrin and hemoglobin levels within the PBS solution. Fig. 4D(III) showed distinct ablation traces and a marked reduction in thrombus volume. Fig. 4D(b) demonstrates the therapeutic contrast of thrombus before and after LAM intervention. However, Fig. 4E (II, III) also indicated that the corresponding reaction process generated substantial smoke, flares, and combustion phenomena. This presented significant safety concerns for clinical application, posing potential hazards and risks of severe injury. The presented LAM@oil has been demonstrated to be an effective solution to this operational safety issue. The LAM@oil was found to be successful in achieving complete reaction within the PBS solution within 12 minutes (Fig. 4D(IV)), with the majority of the clot at the vial bottom dissolving almost entirely within 4 minutes (Fig. 4D(V)). As illustrated in Fig. 4D(c), a comparative visualization of the thrombus before and after treatment with LAM@oil was presented, demonstrating its effective thrombolytic capability.

To investigate the thrombolytic capabilities of LAM@oil with different volumes, we established a simulated vascular system. This dynamic thrombus model comprises a constant-flow pump and a three-dimensional (3D) vascular model. Fig. 4E provided a schematic illustration of this dynamic thrombolysis model, where the constant-flow pump generated circulatory fluid flow, and the thrombus was positioned within the 3D vascular. Fig. 4F depicted the dynamic thrombolysis process within 1 second of contact, the LAM@oil reacted with water molecules, generating bubbles. After 2 minutes, the thrombus underwent color changes due to thermal and alkaline ablation, accompanied by the release of dissolved dark thrombus fragments, which were carried away by fluid flow. Fig. 4G presented the residual thrombus weight curves following treatment with different samples. Due to the shear forces exerted by the circulating simulated blood, the residual weight of thrombi treated with PBS decreased to 80.3%. Thrombi treated with 10 μL LAM@oil exhibited a residual weight reduction to 37.4%, while those treated with 20 μL LAM@oil decreased to 29.8%. This finding demonstrated that the thrombolysis rate exhibited a dose-dependent increase with higher LAM dosage. Fig. 4H revealed a significant reduction in clot volume, with post-thrombolysis clots displaying blackened ablation traces resulting from the chemical ablation by hydroxide ions generated during the reaction. The experimental results indicated that LAM@oil exhibits high thrombolytic efficiency, attributable to the synergistic effect between mechanical disruption via bubble-induced micro-explosions and dissolution through chemical-thermal ablation.

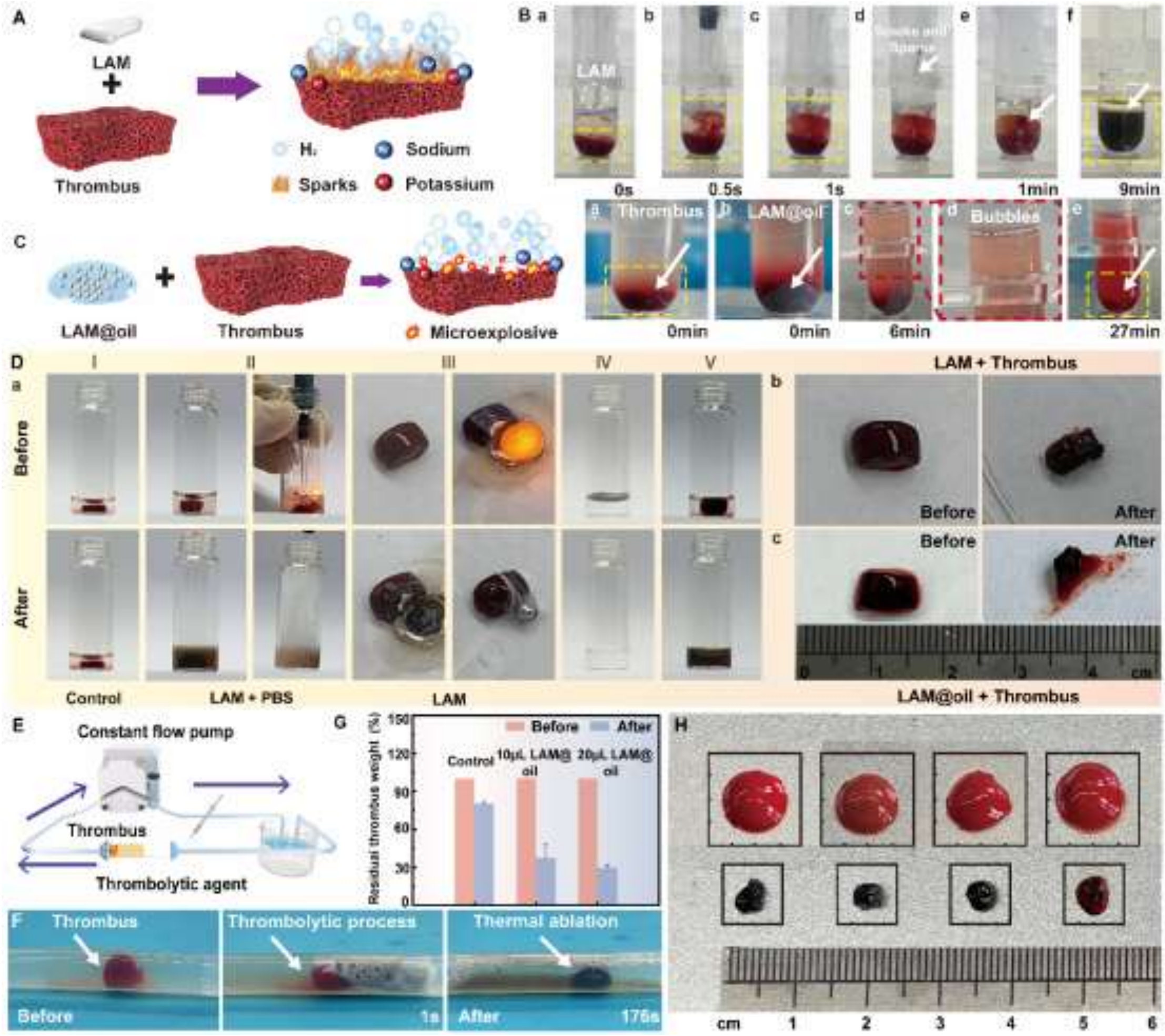

**Fig. 4 In vitro thrombolysis evaluation.** A, Schematic of LAM reacting with thrombus. B, LAM reacting with thrombus in PBS. C, Reaction dynamics of LAM@oil with thrombus and resultant dissolution. D, a. Typical thrombolytic images across experimental groups. I. Control (thrombus in PBS solution), II. LAM injected into the thrombus in PBS solution, and were observed during vigorous reaction, III. LAM was injected into the thrombus under ambient conditions, and combustion phenomena appeared during the reaction. IV. LAM@oil in PBS solution, V. LAM@oil injected into thrombus in PBS solution. b. Thrombus pre-/post-treatment with LAM(D, Group aIII). c. Thrombus pre-/post-treatment with LAM@oil (D, Group aV). E, Hemodynamic flow model with 3D thrombus analogue. F, Real-time in vitro thrombolysis process. G, Residual thrombus mass profiles post-treatment for different groups: control, 10 μL LAM@oil, 20 μL LAM@oil. Data are presented as mean ± SD (n = 4, independent experiments). H, Thrombus analogue comparison pre- versus post-treatment with LAM@oil in the hemodynamic flow model.

## 2.4 Evaluation of *in vivo* arterial thrombolysis efficacy

To assess the in vivo thrombolytic efficacy of LAM@oil, a carotid artery thrombus model was established in Sprague-Dawley (SD) rats. Fig. 5A illustrated the experimental setup used for *in vivo* thrombolysis observation. In healthy SD rats, the carotid artery was characterized by patent blood flow and a bright red appearance (Fig. 5B(a) & (b)). Arterial thrombosis, induced using the well-established $FeCl_3$ injury method, resulted in near-complete blood flow occlusion, significantly reducing flow velocity to negligible levels and conferring a dark reddish hue to the vessel (Fig. 5B(c) & (d)).

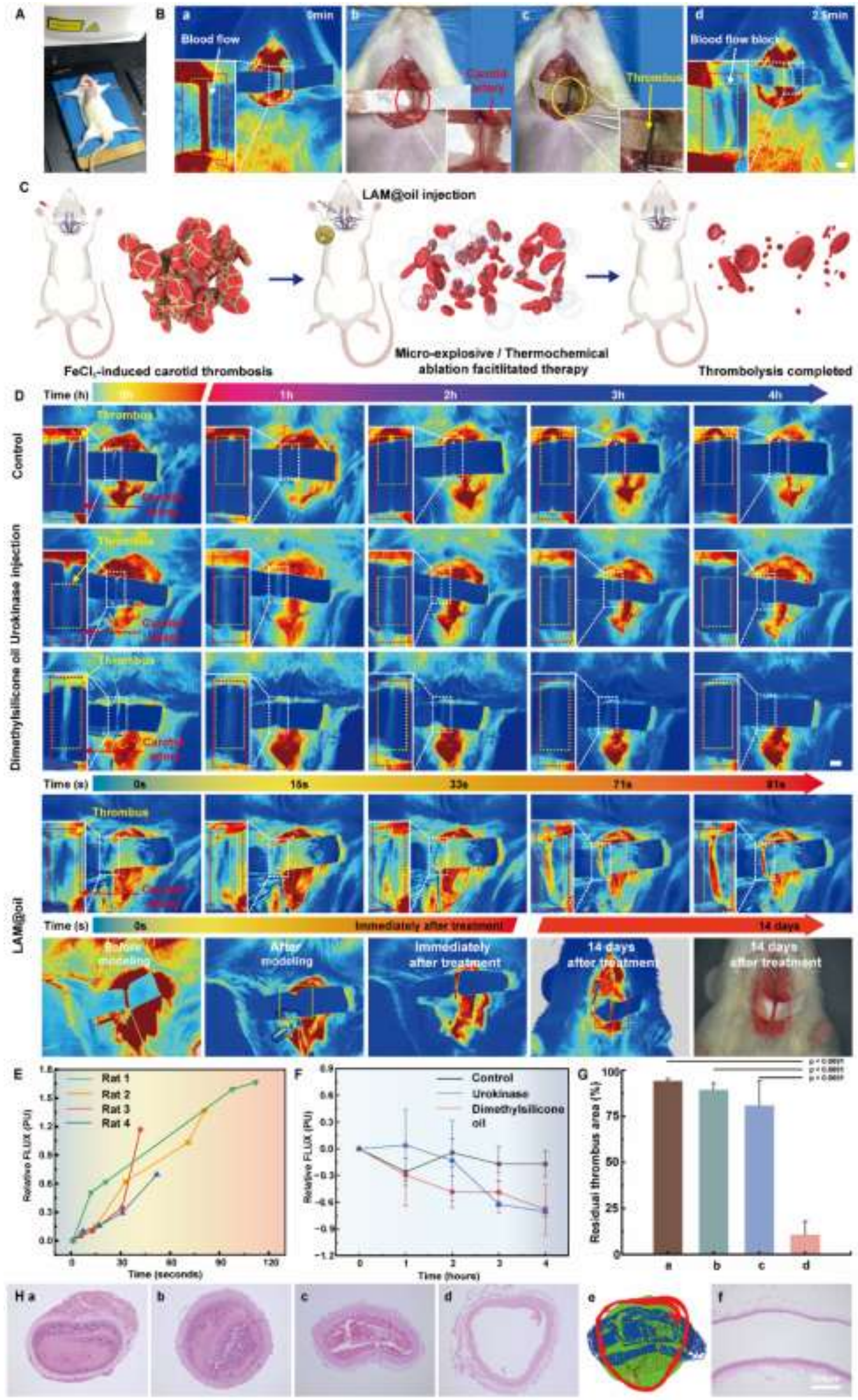

**Fig. 5 In vivo arterial thrombolysis evaluation**. A, Experimental setup for LAM@oil mediated thrombolytic therapy in SD rat carotid thrombosis. B, Blood flow monitored by the LSBFMS (Laser Speckle Blood Flow Monitoring System) pre-thrombosis (0 min) and post-thrombosis (2.5 min).

Scale bar: 10 mm. C, Schematic of micro-explosive/thermochemical synergistic thrombolysis in $FeCl_3$-induced carotid thrombosis in SD rat. D, LSBFMS-monitored blood flow in the carotid arteries of the SD rats: Control (PBS solution), UK, dimethylsilicone, and 10 µL LAM@200µL oil. Scale bar: 10 mm. E, Blood perfusion over time for LAM@oil group (n = 4). F, Blood perfusion over time for control (PBS solution), UK, and dimethylsilicone groups (n = 4 per group). G, Corresponding residual thrombus area (H Group a-d). Data are presented as mean ± SD (n = 4, independent experiments). One-way ANOVA with Tukey's multiple comparisons test was used for the analysis of data. ****$p < 0.0001$. H, Representative H&E staining photomicrographs of the carotid thrombosis after treatment. a: Control (PBS solution), b: Dimethylsilicone oil, c: UK, d: 10 µL of 10 µL LAM@200µL oil (32.26 ± 1.11 µL/kg), e: Comparative analysis of residual thrombosis across experimental groups. f: Longitudinal section of blood vessels in Group d. Scale bar: 300 µm.

To evaluate the *in vivo* thrombolytic efficacy of LAM@oil, the agent was administered distal to the site of carotid artery thrombosis in rats via intravascular injection (Fig. 5C). In the LAM@oil group, enhancement of blood flow signals was observed within a short duration (15 s), attributable to the synergistic interplay of microbubble cavitation and thermo-chemical ablation. At approximately 33 seconds, a progressive increase in thrombolytic efficacy was observed, coinciding with the microbubble cavitation reaction, accompanied by a gradual intensification of blood flow signals. Successful recanalization, as evidenced by robust blood flow signals, was achieved within 81 s. We further evaluated the long-term durability of the therapeutic efficacy of LAM@oil. Immediately after thrombus model establishment, LAM@oil intervention enabled rapid vascular recanalization within a short period, followed by a 14-day long-term follow-up observation on the treated SD rats. The results demonstrated that at 14 days post LAM@oil treatment, the vascular lumens of the rats still exhibited continuous, uniform, and high-brightness blood flow signals, which were consistent with the hemodynamic characteristics of healthy carotid arteries (Fig. 5D). No intermittent blood flow signals, hypoperfused dark areas, or luminal stenosis were detected in the imaging regions, directly confirming that this therapeutic strategy can effectively eliminate the risks of long-term recurrent thrombosis and vascular restenosis. The time-dependent therapeutic profiles across different treatment groups were depicted in Fig. 5D. Thrombolytic trends exhibited variability across groups during the 4-hour observation period. In the saline control group (control), persistent thrombotic occlusion yielded only minor blood flow signals. Urokinase (UK), a clinically established thrombolytic agent, elicited a modest enhancement in blood flow signals within the first hour. Blood flow signals progressively diminished over the subsequent 3 hours, indicating insufficient thrombolysis followed by recurrent thrombosis. The dimethyl silicone group exhibited progressively weakening blood flow signals throughout the 4-hour treatment. The deterioration was attributed to the inherent inability of dimethyl silicone alone to mediate thrombus dissolution, resulting in progressive accumulation of blood cells at the original thrombus site, thereby exacerbating vascular occlusion. The comparative analysis of blood flow restoration across experimental groups visually demonstrated that the LAM@oil group exhibited significantly superior hemodynamic recovery compared to all other treatment groups. Quantitative analysis of blood flow across the experimental groups was presented in Fig. 5E and 5F. Except for the LAM@oil group, all other groups demonstrated a consistent decline in blood flow over time. Conversely, the LAM@oil group demonstrated progressively amplified blood flow enhancement, achieving pronounced thrombolytic efficacy within 90 seconds. The rapid onset of action was closely aligned with our in vitro thrombolysis results. The administration of 10 μL of 10 μL LAM@200 μL oil (32.26 ± 1.11 μL/kg) confirmed that a minute quantity of the agent was capable of eliciting significant thrombus dissolution in vivo within a short timeframe. The UK group showed initial therapeutic efficacy, reflected by increased blood flow within the first hour. However, given the UK's short half-life (typically ~20 minutes) and its systemic clearance duration (ranging from 30 minutes to 2 hours, subject to individual variation), blood flow progressively diminished over the subsequent 3 hours, concomitant with declining drug concentration. Among the remaining groups, the dimethyl silicone group manifested a more substantial reduction in blood flow compared to the saline control, suggesting simethyl silicone oil alone may paradoxically exacerbate thrombotic occlusion through progressive intravascular accumulation. The observation further substantiated that the successful *in vivo* thrombolysis mediated by LAM@oil was principally attributable to its active constituent, NaK. H&E staining of thrombi further corroborated the thrombolytic effect of the NaK. Fig. 5H presented the residual thrombus areas across groups following distinct treatments. Fig. 5H(e) visually demonstrated that the present group exhibited the smallest residual thrombus cross-sectional area among all experimental groups, underscoring its superior therapeutic efficacy. After a 4-hour treatment period, the relative thrombus cross-sectional areas for the saline control, dimethyl silicone, and UK groups were approximately 94.15% ± 1.37%, 89.31% ± 3.72%, and 80.86% ± 13.32%, respectively. In contrast, the relative thrombus area in the LAM@oil treated group was markedly reduced to approximately 10.87% ± 7.16% (Fig. 5G).

Collectively, the *in vivo* thrombolysis findings demonstrated that the LAM@oil exhibited markedly accelerated thrombus dissolution velocity and superior thrombolytic efficiency relative to the other three experimental groups. This enabled the achievement of both speed and efficacy for rapid and highly efficient thrombolysis within an exceptionally short timeframe. Hematoxylin-eosin (H&E) staining was performed on the major organs of rats that achieved immediate thrombolysis following LAM@oil treatment, and no histopathological toxicity was observed. These findings confirm the favorable biocompatibility and safety profile of LAM@oil. Furthermore, H&E staining of major organs from non-thrombotic rats following 14-day intravascular administration of LAM@oil revealed no adverse histopathological alterations. Concurrently, H&E analysis of vascular sections harvested from the arterial reaction zones demonstrated preserved healthy tissue architecture, further corroborating that the LAM@oil induced no significant collateral damage to adjacent vasculature. We also conducted experimental tests on the long-term safety and survival of rats undergoing *in vivo* thrombolytic therapy with LAM@oil. During the 14-day follow-up period, hematological examinations and histopathological analyses of major organs (heart, liver, spleen, lung, and kidney) all fell within the normal range. The survival rate of rats was 100% at day 14, with stable body weight and no significant decrease during the observation period. These rats maintained normal behaviors such as eating, grooming, and movement. Notably, rats that underwent thrombosis modeling but did not receive LAM@oil treatment died within 6 days after modeling. These data not only confirm that carotid artery thrombosis occlusion leads to animal death without intervention but also indirectly verify that our thrombolytic method can effectively recanalize occluded blood vessels and save animal lives without causing systemic toxicity or adverse effects on animal health. Such highly localized thermo-chemical processes ensured operational safety while facilitating thrombus ablation, as thermal energy and hydroxide are released exclusively at the target site. To further explore the translational potential of LAM@oil, we established a rabbit femoral artery thrombosis model and conducted *in vivo* thrombolysis experiments using LAM@oil on this model. LSBFMS clearly captured continuous and uniform blood flow signals in the treated artery of the rabbit after LAM@oil administration, directly confirming the successful achievement of rapid thrombolysis and vascular recanalization. These findings verify that our method possesses core operational feasibility and favorable thrombolytic efficacy in larger animal models with hemodynamic characteristics more analogous to humans, laying a critical foundation for advancing the clinical translation of LAM@oil to benefit patients with thrombotic vascular diseases.

## 2.5 Biological safety of LAM@oil

Most biocompatibility evaluations of samples rely on histological analysis. Since the sample is injected into the circulatory system, histological analysis alone is insufficient to comprehensively assess the safety of the injectable thrombolytic agent LAM@oil in the blood. Therefore, we supplemented blood routine analysis and serum biochemical test. Sprague-Dawley (SD) rats were randomly divided into two groups: the control group (saline injection) and the LAM@oil treatment group. At 2 days (short-term) and 14 days (long-term) post-injection, rats were euthanized, and whole blood samples were collected for subsequent analyses.

Fig. 6A showed the blood test results of rats treated with the LAM@oil over different time periods, indicating that the sample caused no toxicity to the liver or kidney organs in the rats. Compared with the control group, rats injected with LAM@oil showed no significant changes in hepatotoxicity markers including (aspartate aminotransferase (AST)), alanine aminotransferase (ALT), total protein (TP), albumin (ALB), total bilirubin (TBIL), globulin (GLB) or nephrotoxicity markers [urea and creatinine (CREA) at both short-term (2 days) and long-term (14 days) intervals. Additionally, we performed a complete blood routine analysis to supplement the hematological analysis. Key erythrocyte parametersincluding red blood cell count (RBC), mean corpuscular volume (MCV), mean corpuscular hemoglobin (MCH), hemoglobin (HGB) and leukocyte-related parameters neutrophil percentage (NEU%), monocyte percentage (MON%) showed no significant changes between rats treated with LAM@oil and the control group at both short-term (2 days) and long-term (14 days) intervals (Tables S3-S5). The majority of hematological parameters obtained were within normal reference ranges, demonstrating that the injected LAM@oil did not induce toxicity in rats during either short-term or long-term treatment periods. Integrating the above experimental results, the LAM@oil exhibited low hematotoxicity in vivo. Fig. 6B illustrated the distribution of sodium and potassium ion concentrations in major organs (heart, liver, spleen, lungs, and kidneys) of rats 14 days post-treatment with LAM@oil, with all parameters remaining within normal physiological ranges. Notably, the byproducts generated during LAM@oil mediated thrombolysis are sodium and potassium ions, which are essential physiological electrolytes beneficial to the human body. The sodium and potassium ions are biocompatible and ultimately excreted through natural bodily pathways. Fig. 6C showed the H&E staining results of major organs (heart, liver, spleen, lungs, and kidneys) from rats treated with LAM@oil over different time periods, revealing no significant pathological alterations or inflammatory lesions at either short-term (2 days) or long-term (14 days) intervals, indicating that the LAM@oil caused no histotoxic effects on the primary organs of the rats. Throughout the observation period post-injection, all rats exhibited normal behavioral patterns with no observed inflammatory responses, necrosis, or mortality.

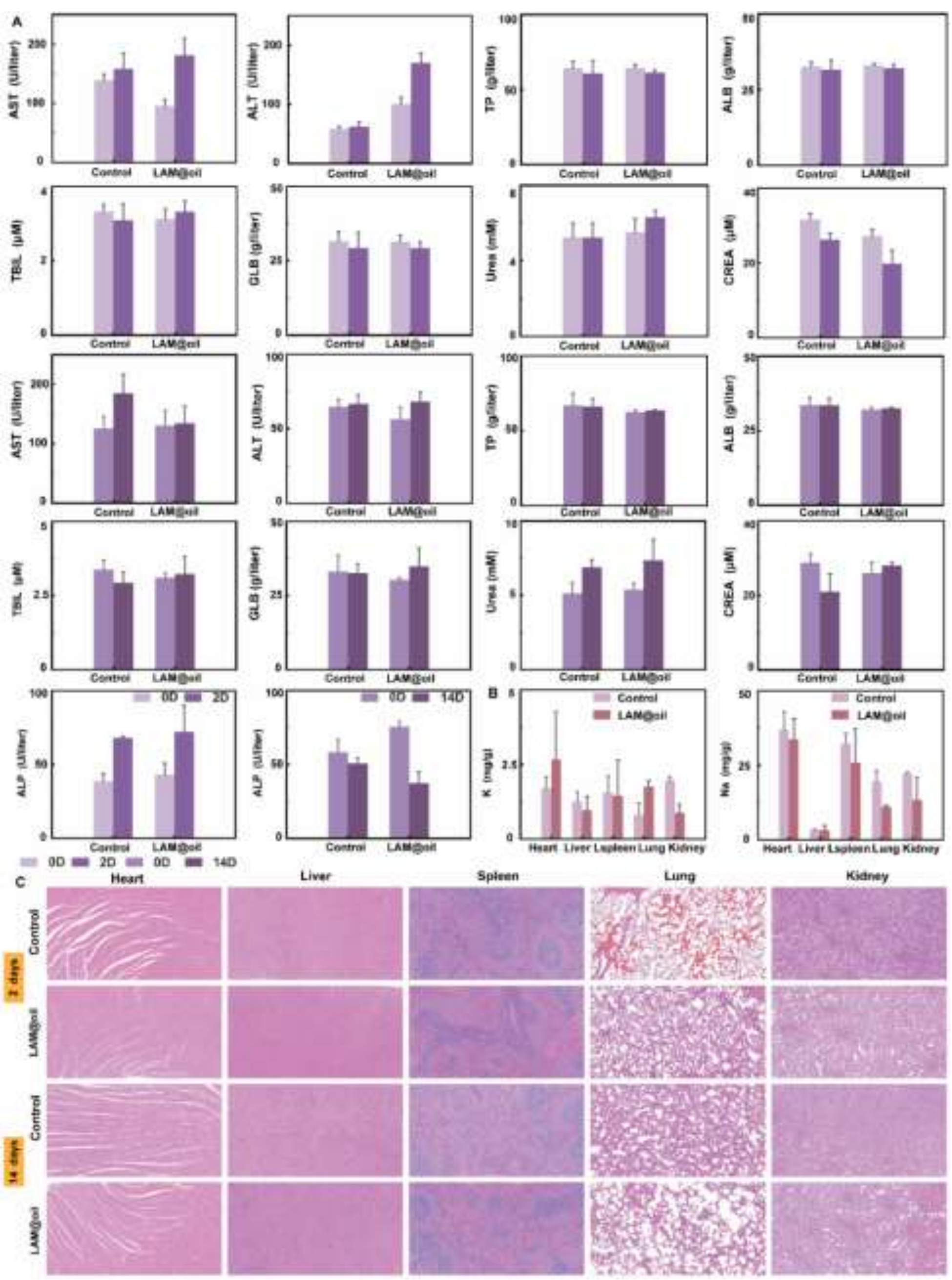

**Fig. 6 In vivo safety assessment of LAM@oil.** A, Blood biochemistry analysis (2-day and 14-day) in rats treated with LAM@oil. Hepatotoxicity markers: aspartate aminotransferase (AST), alanine aminotransferase (ALT), total protein (TP), albumin (ALB), total bilirubin (TBIL), globulin (GLB), alkaline phosphatase (ALP); Nephrotoxicity markers: urea (Urea) and creatinine (CREA). B, Sodium ($Na^+$) and potassium ($K^+$) levels in major organs (heart, liver, spleen, lung, and kidney) of rats treated with LAM@oil for 14 days. C, Representative H&E-stained histological sections of major organs (heart, liver, spleen, lung, and kidney) from the control group, mice treated with LAM@oil for 2 days or 14 days.

With the passage of time, the silicon content in the blood decreased significantly and remained at an extremely low level over the long term. The minimal residual silicon showed no accumulation tendency, which is consistent with its well-established safety profile in clinical practice. These results confirm that the dimethylsilicone used in our study does not pose a risk of accumulation in the circulatory system. Additionally, minor blood clots potentially generated during treatment were predominantly degraded by hemodynamic forces into microparticles that were subsequently cleared via the anti-occlusion system of the glomerular filtration membrane. In addition, we would like to clarify that under the therapeutic conditions of this study, the amount of hydrogen gas generated is extremely small due to the low therapeutic dose of LAM@oil. Furthermore, hydrogen gas is locally produced at the thrombus site in a slow and controllable manner (mediated by the oil-phase barrier), which avoids rapid accumulation. However, it should be noted that if high-dose treatment is required for other scenarios in subsequent applications, attention should be paid to the risk of hydrogen embolism.

Comprehensive experimental evaluation confirmed that the injected LAM@oil did not induce toxicity in rats during the treatment period or subsequent days, while exhibiting low hematotoxicity, with its reaction byproducts being biocompatible, safe, and metabolizable, collectively demonstrating favorable biocompatibility.

## 3. Discussion

The micro-explosive thermochemical thrombolysis (METCT) strategy mediated by injectable liquid alkali metal (LAM@oil) represents a transformative and urgently needed advancement in the treatment of acute thrombotic diseases. Unlike conventional pharmacological thrombolytics, which are hampered by systemic bleeding risks, short half-lives, and delayed recanalization or invasive mechanical thrombectomy that requires specialized equipment and prolonged operation, the LAM@oil system offers an unprecedented combination of rapidity, efficacy, and simplicity. This approach directly addresses a critical societal and clinical imperative: the need for a safe, fast, and accessible thrombolytic intervention capable of mitigating ischemic damage within the narrow therapeutic window of conditions such as acute myocardial infarction and stroke. The procedure is remarkably straightforward, requiring only a single injection to initiate near-instantaneous (within 90 seconds), multimodal thrombus dissolution via synergistic micro-explosive, alkaline, and thermal mechanisms. This capability for ultra-rapid recanalization positions LAM@oil as a potentially life-saving intervention in time-sensitive emergencies where every second counts.

Compared to existing strategies, LAM@oil demonstrated superior thrombolytic performance through three synergistic pathways: (i) The generation of instantaneous cavitation effects by the continuous implosion of hydrogen bubbles at elevated temperatures produced high-energy microjets. These microjets mechanically disrupted thrombus macrostructures through tearing. (ii) Intense exothermic redox reactions between LAM and thrombus components created alkaline environments, catalyzing proteolytic hydrolysis. (iii) Thermal denaturation of the fibrin network at elevated temperatures achieved thermothrombolysis. The thrombolytic velocity and efficacy exhibited positive correlation with increasing LAM concentration in the medium, featuring self-limiting thermal properties (temperature regulated by reactant dosage) that enabled adjustable formulation ratios based on clinical requirements. Moreover, LAM@oil demonstrates superior performance across multiple metrics: where urokinase left ~81% residual thrombus in our in vivo model, LAM@oil reduced it to ~11%, without inducing systemic bleeding or requiring complex infusion systems. Unlike nanomaterial-based agents that struggle with penetration, stability, and biocompatibility, LAM@oil leverages a controlled, localized reaction that is self-limiting and leaves only benign, physiological electrolytes ($Na^+$, $K^+$) as byproducts. Furthermore, the encapsulation strategy tames the inherent reactivity of alkali metals, enabling safe administration and eliminating risks of unintended combustion or vessel injury.

Critically, the comprehensive safety profile of LAM@oil is unequivocally demonstrated through rigorous in vivo assessments, including histopathology, hematological parameters, and ion distribution studies, all confirming an absence of toxicity, inflammation, or organ damage. The combination of rapid and complete thrombolysis with this exceptional safety margin renders LAM@oil not merely an incremental improvement, but a paradigm shift in thrombosis management, which is green, economical, and immediately applicable. This innovative platform meets the urgent demand for reliable and efficient thrombolysis and holds significant promise for clinical translation.

## 4. Conclusion

In summary, we have established an injectable LAM@oil system with the mechanism of micro-explosive and thermochemical thrombolysis (METCT). The proposed LAM@oil represents a green and synergistic thrombolysis strategy that is injectable, straightforward, economical, and practical. The encapsulation of LAM with dimethylsilicone has been shown to reduce its high reactivity and enhance stability, avoiding violent explosive combustion. In addition, the generation of microbubbles by controlled aqueous reactions was demonstrated to enable safer thrombus ablation. In vitro thrombolytic assessments in static and dynamic flow models were conducted, showing the trace amounts of LAM@oil's capability in thrombi debulking, which can be confirmed by the reduced thrombus weight to 29.8% of the initial mass. In vivo carotid artery thrombosis models in rats showed that 10 μL of LAM@200 μL oil achieved rapid and near-complete thrombolysis within a remarkably short 90 seconds, demonstrating its potential as a life-saving intervention. This speed and procedural simplicity-requiring only a single, straightforward injection-stand in stark contrast to the complexity and time delays associated with conventional pharmacologic thrombolysis (often requiring prolonged infusions and carrying bleeding risks) or mechanical thrombectomy (demanding specialized equipment and highly trained operators). Furthermore, the post-thrombolysis byproducts of LAM@oil mainly included sodium and potassium ions, which are intrinsic physiological electrolytes and biocompatible, bioavailable, and entirely non-toxic. These byproducts underwent rapid biodegradation and complete systemic elimination. Crucially, our comprehensive safety assessment confirms that LAM@oil presents an exceptionally low-risk profile. The controlled reaction mechanism, biocompatible byproducts, and rapid systemic clearance collectively ensure complete operational and biological safety, leaving no residual risk or toxicity concerns. The potent efficacy and this substantial biosafety profile indicated significant clinical potential for thrombotic disorders. We envision the LAM@oil system as the next-generation thrombolytic technique, offering a transformative paradigm shift to significantly improve the thrombolytic efficacy, speed, accessibility, and safety for clinical non-pharmacological thrombosis management, directly addressing the urgent societal demand for better, faster, and safer solutions to thrombotic emergencies.

## 5. Experimental Section

### Materials

Liquid sodium-potassium alloy (NaK), Dimethylsilicone (D806807,Macklin Biochemical Co.,

Ltd., Shanghai, China), PBS solution (Hyclone, USA), 10% fetal bovine serum (FBS, Hyclone, USA), DMEM culture medium (Hyclone, USA), 20% ferric chloride ($FeCl_3$, Shanghai Aladdin Biochemical Technology Co., Ltd., China), Handheld ultrasonic homogenizer (Branson SFX55, EMERSON, USA), Thermocouple probe (KAIPUSEN, Suma Electrical Instrument Co., Ltd., China), Gas chromatograph (GC-2014C, Shimadzu Co., Ltd., Japan), Computational modeling software (Comsol Multiphysics), Peristaltic pump (Aurora100,MainLand Electronic Technology (Baoding) Co., Ltd., China), Laser speckle blood flow monitoring system (SIM BFI-ZOOM, Wuhan Xunwei Optoelectronic Technology Co., Ltd., China), Inductively coupled plasma optical emission spectrometer (ICP-OES, iCAP 7400, Thermos, Waltham, USA), CCK-8 assay kit (Solarbio, CA1210), Sprague-Dawley (SD) rats (Vital River Laboratory Animal Technology Co., Ltd., Beijing, China), Deionized water were used for all experimental procedures. EC cell line (Research Resource Identifier (RRID): CVCL_2959, purchased from Wuhan Servicebio Co., Ltd., Wuhan, China. The cell line was authenticated and tested to be free of mycoplasma and other microbial contamination prior to experimental use).

**Fabrication of LAM@oil**

LAM@oil was made from liquid sodium-potassium alloy (NaK) and dimethylsilicone. The LAM exhibits fluidity in its liquid state at room temperature. Precisely measured volumes of NaK (5 μL, 10 μL, 20 μL) and dimethylsilicone (20 μL, 40 μL, 80 μL, 100 μL) were carefully added to the bottom of 1.5 mL EP tubes according to experimental requirements, followed by thorough mixing via sonication with an ultrasonic probe (Handheld ultrasonic homogenizer). At room temperature, using parameters of 2-second pulses at 10-second intervals for 5 cycles at 150 W maximum output to ensure uniform emulsification. The resulting stable homogeneous LAM@oil dispersion underwent UV disinfection prior to subsequent experiments.

**In vitro temperature profiling and hydrogen evolution assessment**

LAM@oil was prepared at varying NaK concentrations (5 μL NaK -20, 40, 80 μL oil, 10 μL NaK-20, 40, 80 μL oil, 20 μL NaK-40, 80, 100 μL oil), with solution temperatures recorded continuously for 190 seconds using a thermocouple. For hydrogen quantification, prepared samples in 12×75 mm tubes containing 1 mL PBS were sealed with rubber stoppers and parafilm, then incubated at 37 °C for 3 hours unless otherwise specified. The gas products were quantified by withdrawing gas samples from the reactor and injecting them into a gas chromatograph (GC-2014C, Shimadzu Co., Ltd., Japan) equipped with an active-carbon-packed column and a thermal conductivity detector (TCD).

**Simulation of the LAM@oil mediated heat and mass transfer model**

The heat and mass transfer computational model for thermal ablation of vascular thrombus involves multiple aspects such as heat transfer, mass transfer and coupling effects, and needs to simultaneously consider the influence of thermal and chemical effects on the thrombolytic effect. The temperature and ion concentration distribution described in Fig. 3 were solved in COMSOL Multiphysics. The classical Pennes biological heat transfer model was adopted, assuming that the heat produced by blood and metabolism was uniformly distributed.

Heat transfer mainly includes heat conduction, and heat conduction describes the transfer of heat in thrombus and vascular tissue with the equation:

$$\rho c_p \frac{\partial T}{\partial t} = \nabla \cdot (k \nabla T) + Q_m + Q_b + Q \tag{2}$$

where $\rho$ is the density, $c_p$ is the specific heat capacity, $k$ is the thermal conductivity, $T$ is the temperature, $Q_m$ is the heat production due to metabolism, $Q_b$ is the heat production due to blood perfusion in biological tissues, and $Q$ is the heat released from chemical reactions.

The transfer of mass mainly consists of chemical reactions and diffusion of oil-coated LAM composite systems in the thrombus and blood vessels, and the rate equation describing the reaction of the systems with the thrombus is:

$$R=k_fC \tag{3}$$

where $K_f$ is the reaction rate constant.

The diffusion equation for the composite system in the thrombus and blood vessels is:

$$\frac{\partial C}{\partial t} = D\nabla^2 C - R \tag{4}$$

where $C$ is the concentration, $D$ is the diffusion coefficient, and $R$ is the chemical reaction rate.

**Artificial thrombus preparation**

The in vitro thrombus model was prepared based on previously reported literature[47,48]. Artificial thrombi were prepared by collecting 5 mL fresh whole blood, aliquoting into 200 µL volumes per tube, incubating tubes at room temperature for 5 hours, and then transferring to 4 °C for 1-2 days to achieve maximal clot retraction and stabilization.

**In vitro static/dynamic thrombolysis evaluation**

For static thrombolysis assessment: LAM was added to tubes containing 1 mL PBS and 200 µL thrombus model. The tubes were then incubated at 37 °C, and the reaction was documented. Subsequently, the thrombus was extracted for gravimetric analysis at predetermined intervals. The analysis was performed using trace amounts (1 µL) of non-encapsulated LAM. A series of experiments was conducted in parallel to assess the efficacy of different solutions (PBS, LAM in PBS, 10 µL LAM@oil in PBS) with 200 µL thrombi under identical conditions, recording morphological changes photographically.

For dynamic thrombolysis modelling: Thrombi were positioned within mock arterial vessels where PBS was peristaltically circulated (Peristaltic pump) to simulate blood flow. After introducing test solutions (10 µL, 20 µL LAM@oil), residual thrombus weights were measured at designated timepoints to calculate thrombolytic efficiency.

**In vivo thrombus modelling and thrombolytic evaluation**

Multiple male SD rats (weight 300 ± 10 g, purchased from Beijing Vital River Laboratory Animal Technology Co., Ltd.) were used with the protocol approved by the Animal Ethics Committee (License No.: IACUC-IPC-24073). All rat experiments complied with the *Guidelines for the Management and Use of Laboratory Animals* issued by the Ministry of Health of the People's Republic of China. SD rats were freely fed and maintained in a constant temperature/humidity environment with 12h light/dark cycles. After a 3-day acclimation period, model construction commenced. We established the in vivo thrombus model (rat carotid artery thrombosis model) according to published methods[49]. Rats were weighed and anesthetized via intramuscular injection of Zoletil + Xylazine (0.1 ml/kg), then fixed in a supine position after successful anesthesia. A vertical incision was made on the neck of SD rats to expose the carotid artery. A filter paper (1.5 × 3 mm, soaked in 20% $FeCl_3$) was wrapped around the carotid artery. After 2.5 minutes, the carotid artery was rinsed with excessive saline. The arterial segment darkened and exhibited slight swelling, forming a carotid thrombus within 5 minutes. Rats were then randomized into 5 groups receiving different samples: control (saline) group, dimethylsilicone group, UK group, and10 μL of 10 μL LAM@200 μL oil (32.26 ± 1.11 μL/kg). Carotid blood flow was monitored pre-/post-thrombosis and during intervention using the Laser Speckle Blood Flow Monitoring System (LSBFMS). Acquired hemodynamic data were analyzed with SIM BFI software to evaluate thrombolytic efficacy. Post-experiment, vessels and thrombi were harvested for imaging and H&E staining.

**Cell culture and viability assay**

EC cells (vascular endothelial cells) purchased were cultured in high-glucose Dulbecco's Modified Eagle Medium (DMEM, Hyclone, USA) supplemented with 10% fetal bovine serum (FBS, Hyclone, USA) and 1% penicillin/streptomycin. The temperature of the cell culture box was set at 37 °C and the $CO_2$ content was 5%. The cells were incubated under these conditions. For viability assessment, cells were seeded at 50,000 cells/mL concentration (100 μL per well) in 96-well plates. After 12 hours, DMEM was replaced with fresh DMEM (100 μL per well) containing different concentrations of LAM@oil (10 μl LAM @20 μL, 40 μL, 80 μL oil, 20 μL, LAM@ 40 μL, 80 μL, 1000 μL oil), while control wells received fresh DMEM (100 μL per well). Following 24-hour incubation, media were removed and replaced with complete DMEM containing 10% CCK-8 (100 μL per well), then measured for absorbance intensity at 450 nm using a microplate reader after 1-2 hours incubation at 37 °C, with cell viability calculated as detailed below.

$$Viability(\%) = (\frac{OD_{drug} - OD_{blank}}{OD_{control} - OD_{drug}}) \quad (5)$$

**In vivo safety evaluation**

After environmental acclimatization, SD rats (weight 300 ± 10 g) were divided into 4 groups [control, LAM@oil (100 μL)] for toxicity testing at day 2 and day 14 (n=4 per group). Subsequently, experimental group rats were euthanized on days 2 and 14, with blood samples collected and major organs harvested for blood routine test, serum enzyme analysis, and histological examination. Simultaneously, sodium and potassium ion levels in major organs were quantified using ICP-OES.

**Statistical Analysis:**

Statistical computations were performed using GraphPad Prism. All data were presented as mean ± SD. All experimental data were compared using one-way ANOVA, followed by Tukey's multiple analysis to determine the level of statistical signifi-cance between two or multiple groups. Significant differences between the groups were indicated by n.s. ($p > 0.05$), *$p < 0.05$, **$p < 0.01$, ***$p < 0.001$, and ****$p < 0.0001$.

**Acknowledgements**

This work is partially supported by the Strategic Priority Research Program of the Chinese Academy of Sciences (No. XDB1030000) and Beijing United Fund (No. L252063).

**Author contributions**

Conceptualization, X.L., J.L.; Methodology, X.L., Y.H., H.Q., W.R. and J.L.; Investigation, X.L.; Formal analysis, X.L., Y.H.; Validation: X.L.; Visualization, X.L., Y.H., B.W., and M.G.; Data Curation, X.L., Y.H., and J.Z.; Funding acquisition, H.Q., W.R., J.L.; Resources, W.R., J.L.; Software, X.L., Y.H, and J.Z.; Project administration, W.R., J.L.; Supervision, W. R., J.L.; Writing—original draft, X.L., Y.H.; Writing—review and editing, X.L., Y.H., W. R., and J. L.